\documentclass{emulateapj}

\def\beq{\begin{equation}}
\def\eeq{\end{equation}}
\def\beqar{\begin{eqnarray}}
\def\eeqar{\end{eqnarray}}

\def\pfrac#1#2{\left( \frac{#1}{#2} \right)}
\def\avg#1{\langle #1 \rangle}

\def\msol{M_\odot}

\def\csfr{\dot{\rho}_\star}

\def\eobs{E}
\def\eem{E_{\rm em}}

\def\omm{\Omega_{\rm M}}
\def\oml{\Omega_{\Lambda}}
\def\mgas{M_{\rm gas}}

\def\halpha{\mbox{H$\alpha$}}
\def\lha{L_{\rm H\alpha}}
\def\tracer{L}

\def\grld{{\cal L}_\gamma}

\def\apjl{Astrophys.~J.~Lett.}
\def\apj{Astrophys.~J.}
\def\aj{Astronomical~J.}
\def\prd{Phys.~Rev.~D}
\def\aap{Astron.~Astrophys.}
\def\mnras{Monthly Not.~Royal Astr.~Soc.}

\def\la{\mathrel{\mathpalette\fun <}}
\def\ga{\mathrel{\mathpalette\fun >}}
\def\fun#1#2{\lower3.6pt\vbox{\baselineskip0pt\lineskip.9pt
 \ialign{$\mathsurround=0pt#1\hfil##\hfil$\crcr#2\crcr\sim\crcr}}}

\shorttitle{Cosmic Gamma-Ray Background}
\shortauthors{Fields et al.}

\begin{document}

\title{Cosmic Gamma-Ray Background 
from Star-Forming Galaxies}

\author{Brian D. Fields}
\affil{Departments of Astronomy and of Physics, 
 University of Illinois, Urbana, IL 61801, USA}
\email{bdfields@illinois.edu}

\author{Vasiliki Pavlidou\altaffilmark{1}}
\affil{California Institute of Technology, Pasadena, CA 91125, USA}

\and

\author{Tijana Prodanovi\'c}
\affil{Department of Physics, University of Novi Sad, 
Trg Dositeja Obradovi\'{c}a 4, 21000 Novi Sad, Serbia}

\altaffiltext{1}{Einstein Fellow}

\begin{abstract}
The origin of the extragalactic gamma-ray background is a pressing
cosmological mystery.  The {\it Fermi} Gamma-Ray Space Telescope has
recently measured the intensity and spectrum of this background; both
are substantially different from previous measurements. We
present a novel calculation of the gamma-ray background from
normal star-forming galaxies.  Contrary to longstanding
expectations, we find that numerous but individually faint normal
galaxies may comprise the bulk of the {\it Fermi} signal, rather than rare but
intrinsically bright active galaxies.  This result has wide-ranging
implications, including: the possibility to probe the cosmic
star-formation history with gamma rays; the ability to infer the
cosmological evolution of cosmic rays and galactic magnetic fields;
and an increased likelihood to identify subdominant components from
rare sources (e.g., dark matter clumps) through their large
anisotropy.
\end{abstract}


\section{Introduction}

The {\em Fermi} Gamma-Ray Space Telescope has 
unveiled the high-energy cosmos with unprecedented clarity
and depth.
The gamma-ray sky has been known \citep[e.g.,][]{hunter,sreek}
to be dominated by diffuse emission from the Galactic plane,
while at high Galactic latitudes a diffuse
extragalactic gamma-ray background (EGB) has an important, and at some energies dominant, contribution.
However, before {\em Fermi}, the processes dominating the diffuse emission from
the Galaxy, especially above 1 GeV, were unclear and highly
debated--cf.~discussion on the {\em GeV excess} reported by the
Energetic Gamma-ray Experiment Telescope (EGRET)
\citep[e.g.,][and references therein]{smr00}.
{\em Fermi} has clarified \citep{FermiMidLat09} that the dominant emission
mechanism is cosmic-ray interactions with interstellar gas,
which leads to gamma rays mostly from pion decay in flight, i.e.,
$p_{\rm cr} + p_{\rm ism} \rightarrow p p \pi^0$
then $\pi^0 \rightarrow \gamma \gamma$ \citep{stecker}.

Moreover, the {\em Fermi} EGB differs
from previous EGRET estimates: 
the intensity is {\em fainter} 
and the spectrum {\em steeper}, consistent with a power law
of spectral index $2.41 \pm 0.05$ and integrated intensity $I(>100{\rm MeV})=(1.03\pm0.17)\times10^{-5}{\rm cm^{-2}s^{-1}sr^{-1}}$
\citep{fermi-EGB}.
Our theoretical understanding of the EGB 
must therefore be substantially revised in light of the
new and smaller {\em Fermi} signal.

Pioneering studies investigating the origin of the EGB first considered 
the collective emission form 
star-forming galaxies (like the Milky Way), 
but found this to give a small EGB signal  \citep{sww,bignami}.
\citet{pf02} first incorporated observations of the cosmic star-formation
rate, 
while ~\citet{pf06} made the first estimates of the pionic
contribution from star-forming galaxies.
In both cases, the predicted intensity
was below the then-measured EGB.
Blazars, the brightest extragalactic sources, have been the favored candidates \citep{ss96}.
However, subsequent estimates of their contribution to the EGB 
\citep[e.g.,][]{mucke,cm98,nt07,dermer07} have consistently fallen short.
{\em Fermi} point-source observations suggest that unresolved blazars contribute at most $\sim 23\%$ of the EGB; and thus the mystery has become acute
\citep{fermicounts}.

Here we present a more realistic
model for the EGB from star-forming galaxies, constructed to use as
much  as possible of our substantially improved multiwavelength observational
understanding of these sources, and isolating the signal
from normal star formation (as opposed to starburst galaxies). 

\section{Formalism}

{\em The EGB intensity} is an integral of the gamma-ray luminosity
density $\grld$ (emissivity) over the line-of-sight to the cosmic horizon
\beq
\label{eq:egb}
\frac{dI}{d\eobs} = \frac{c}{4\pi} \int \grld[(1+z)\eobs,z] \ (1+z)
 \left| \frac{dt}{dz} \right| \ dz
\eeq
where $|dt/dz| = (1+z)^{-1} H(z)^{-1} = H_0^{-1} (1+z)^{-1}
[(1+z)^3\omm + \oml]^{-1/2}$, with $H(z)$ the Hubble function.
We adopt a $\Lambda$CDM cosmology, with 
a Hubble parameter
$H_0 = 71 \ \rm km \ s^{-1} \ Mpc^{-1}$,
and cosmological constant and matter density
parameters $\omm = 0.3$ and $\oml = 0.7$
respectively. 

The cosmological inputs to eq.~(\ref{eq:egb}) are well-determined,
so the EGB entirely hinges on 
the luminosity density $\grld$ and its connection to cosmic star formation.
In this study we construct, for the first time, a gamma-ray luminosity function (distribution of sources by gamma-ray luminosity as a function of redshift) for normal galaxies. The luminosity function 
is used to obtain the EGB intensity, by integrating over luminosities and redshift along photon paths.

{\it The star-forming luminosity density} due to
pionic emission follows from the gamma-ray luminosity per star-forming
galaxy $L_\gamma$.  
\beq
\label{eq:grld}
\grld
= \avg{L_\gamma n_\gamma} \equiv 
  \int_0^{L_{\rm max}} L_\gamma \ n_\gamma(L_\gamma) \ dL_\gamma
\eeq
where
$n_\gamma$ 
is the comoving number density of gamma-ray-luminous  galaxies.
The average is taken over the distribution of star-forming galaxy
properties at redshift $z$, i.e., 
the gamma-ray luminosity function which gives the comoving number $n_\gamma(L_\gamma) \ dL_\gamma$
of star-forming galaxies with luminosity in
the range $(L_\gamma,L_\gamma+dL_\gamma)$;
out to a maximum $L_{\rm max}$ determined below.

{\it A galaxy's pionic gamma-ray flux}
scales with the cosmic-ray flux (projectiles)
and the amount of interstellar {\em gas} (targets) in the galaxy. 
Specifically, a galaxy's rest-frame 
pionic gamma-ray luminosity (photon {\em counts} per
unit time) 
is given by
\beqar
\label{eq:gamma-pionic}
L_\gamma(\eem) 
& = & \int \Gamma_{\pi^0 \rightarrow \gamma\gamma}(\eem)  n_{\rm H} \ dV_{\rm ism} \\
& = & \Gamma_{\pi^0 \rightarrow \gamma\gamma}(\eem)
 \ {\cal N}_{p} \ \propto \ \Phi_p M_{\rm gas}
\eeqar
with $\eem$ the photon energy in the emitting frame. 
Here the pionic gamma-ray production rate per interstellar H-atom
is 
$\Gamma_{\pi^0 \rightarrow \gamma\gamma}(\eem) 
 = \avg{\Phi_p \, d\sigma_{\pi^0 \rightarrow \gamma\gamma}/d\eem} \propto \Phi_p$ 
and is proportional to the galaxy's {\em volume-averaged} cosmic-ray proton flux $\Phi_{p}$
\citep{pohl,persic};
the cross-section $d\sigma_{\pi^0 \rightarrow \gamma\gamma}/d\eem$ is
understood to include effects of pion multiplicity and 
of helium and heavier elements in the cosmic rays and interstellar medium.
The factor ${\cal N}_{p} = \int n_{\rm H} \ dV_{\rm ism}$ gives the number  
of  hydrogen atoms in the galaxy's interstellar medium,
summed over all states--molecular, 
atomic, and ionized.  
This term is 
fixed by the galaxy's total interstellar 
gas mass via ${\cal N}_{\rm H} = X_{\rm H} \mgas/m_p$,
where $X_{\rm H} \approx 0.70$ is the hydrogen mass fraction.

Since cosmic rays are thought to be predominantly accelerated in sites  
associated with massive star formation (supernova remnants, massive  
stellar winds, pulsars, see e.g., \citet{lacki} and references  
therein), their 
flux in eq.~(\ref{eq:gamma-pionic})
should scale with the {\em star formation rate} (SFR) in the galaxy.
Observations by {\em Fermi} and TeV telescopes HESS and VERITAS have  
confirmed this expectation through observations of star-forming  
galaxies \citep{fermi-LMC,sfgal,fermi-1fgl, hess09, veritas09}, for  
which both SFR and total gas mass were independently known.
Thus the cosmic-ray flux is set by the competition of particle acceleration
and losses that
we assume are dominated by escape, as they are in the Milky Way.  
These self-regulating processes tend toward an equilibrium
$\Phi_p \propto \Lambda_{\rm esc} \psi$,
where $\psi$ is the star formation rate and $ \Lambda_{\rm esc}$ is the escape pathlength (assumed constant).
We thus 
 adopt
the scaling 
\beq
\label{eq:scaling}
L_\gamma \propto  M_{\rm gas} \psi
\eeq
of gamma-ray luminosity with a galaxy's gas mass and 
star-formation rate $\psi$ \citep{pf01}.

The scaling in eq.~(\ref{eq:scaling})
represents a central {\em ansatz} regarding gamma-ray production
in Milky-Way-like galaxies whose cosmic-ray losses are dominated by escape.
This physically-motivated relation now has support from 
{\em Fermi} observation of star-forming galaxies for
which the product $M_{\rm gas} \psi$ ranges over several
orders of magnitude.   
A second {\em ansatz} in our model 
is our adoption of a universal cosmic-ray spectrum
of index $s_{\rm cr} = 2.75$.
At energies away from the peak the gamma-ray spectrum
has the same index: $s_{\gamma,\rm had} = s_{\rm cr}$.  
Even though both approximations are certainly simplifying, they represent important benchmark
cases, against which more sophisticated models can be tested.

The scaling law of eq.~(\ref{eq:scaling}) allows us to 
determine the gamma-ray output of any star-forming galaxy,
but only once we normalize our results to a system
in which the cosmic-ray and/or gamma-ray properties are known.
We choose to normalize to the Milky Way, where
the local cosmic-ray flux is well-measured, and the
global star-formation rate is also known.
We assume that
the ratio of cosmic-ray flux to star-formation
rate should be constant for all normal galaxies, i.e., that
\beq
\frac{\Gamma_{\pi^0 \rightarrow \gamma\gamma}}{\Gamma_{\pi^0 \rightarrow \gamma\gamma}^{\rm MW}}
 = \frac{\Phi_{\rm cr}}{\Phi_{\rm cr,MW}} 
 = \frac{R_{\rm SN}}{R_{\rm SN,MW}}
   = \frac{\psi}{\psi_{\rm MW}}
\label{eq:ratio}
\eeq
This scaling encodes the longstanding notion that supernova
remnants accelerate hadronic cosmic rays
\citep[leading to pionic emission, e.g.,][]{fermi-W51C}.

We thus find that for normal galaxies,
\beqar
\label{eq:gammalum}
L_\gamma(\mgas,\psi) 
 &=& X_H \Gamma_{\pi^0 \rightarrow \gamma\gamma}^{\rm MW}
  \frac{\mgas}{m_p} \frac{\psi}{\psi_{\rm MW}}  \\
 &=& 1.7 \times 10^{42} \ {\rm s^{-1}} \
  \ \pfrac{\mgas}{10^{10} \msol} 
  \ \pfrac{\psi}{1 \msol \ \rm yr^{-1}} \ \ .
\nonumber
\eeqar
We adopt the photon emission per hydrogen atom
derived from {\em Fermi} diffuse Galactic observations 
at medium latitudes for photons $> 100$ MeV:
$\Gamma_{\pi \rightarrow \gamma\gamma}^{\rm MW} = 2.0 \times 10^{-25} \
          \rm s^{-1} \mbox{H-atom}^{-1}$
\citep{FermiMW09}.
This value represents a large-scale ($\sim$kpc) spatial averaging of
Galactic cosmic-ray properties, which is appropriate for our global
calculation. 

We can now write the gamma-ray luminosity density for normal galaxies 
\beq
\label{eq:lumdens}
\grld(\eem,z)   = X_{\rm H} \ 
    \frac{\Gamma_{\pi^0 \rightarrow \gamma\gamma}^{\rm MW}(\eem)}
         {\psi_{\rm MW}} \
 \frac{\avg{\mgas(z)}}{m_p} \ \csfr(z)
\eeq
in terms of the cosmic star-formation rate $\csfr(z) = \avg{\psi n_{\rm gal}}$.
We define a mean interstellar gas mass as
\beq
\label{eq:mavg}
\avg{\mgas(z)} \equiv 
  \frac{\avg{\mgas \psi n_{\rm gal}}}{\avg{\psi n_{\rm gal}}}
  = \frac{\avg{\mgas \psi n_{\rm gal}}}{\csfr}
\eeq

The assumption that losses are escape-dominated and uniform across galaxies
can only be approximately valid at best.
This is a major uncertainty in our model,  
which will benefit from future data on the EGB and resolved galaxies.
For example, even the leaky-box model can generalize
eq.~(\ref{eq:scaling}) to
$L_\gamma \propto \Lambda_{\rm esc} \psi \mgas$;
variations in the energy dependence of the escape length
$\Lambda_{\rm esc}$ would also change the cosmic-ray and photon
spectral indices which we take as universal.

Indeed, {\em Fermi} observations of the Large Magellanic Cloud
suggest that cosmic-ray confinement and propagation are non-trivial
\citep{fermi-LMC}.
Moreover, starburst galaxies show very high cosmic-ray intensities
within small volumes where inelastic collisions compete with, and sometimes dominate,
outflows to regulate cosmic-ray losses
\citep{pag1996,lacki,torres,thompson,persic,stecker07}.
For this reason, 
eq.~(\ref{eq:gammalum}) provides a rough description of normal escape-dominated
galaxies only; we do not expect it to hold for 
starburst galaxies, which we will exclude below.

\subsection{Gas Mass and Star-Formation Rate}

We can infer a galaxy's interstellar gas mass at a given star-formation rate, via the 
well-established Kennicutt-Schmidt law \citep{schmidt,kennicutt}.
The {\em surface densities} for
star formation and gas are found to be correlated via
$\dot{\Sigma}_{\star}/\msol \rm yr^{-1} kpc^{-2} = 
(2.5 \pm 0.7) \times 10^{-4} 
 (\Sigma_{\rm gas}/\msol {\rm pc}^{-2})^{\mathnormal x}$
with $x = 1.4 \pm 0.15$.
Both normal and starburst galaxies follow this correlation, 
but normal galaxies populate
$\dot{\Sigma}_{\star,\rm normal} \la \dot{\Sigma}_{\star,\rm normal}^{\rm max} \equiv 0.4 \ \msol\  \rm yr^{-1} kpc^{-2}$
while starbursts occupy the opposite regime \citep{kennicutt}. 

To recover a relationship between the global galactic star-formation rate
$\psi = \pi r_{\rm disk}^2 \dot{\Sigma}_{\star}$ and gas mass
$\mgas = \pi  r_{\rm disk}^2 \Sigma_{\rm gas}$,
requires a galactic disks scale length $r_{\rm disk}$.
We take $ r_{\rm disk} = 18.9 \ {\rm kpc}/(1+z)$; observations \citep{erb06}
indicate that this choice
is uncorrelated with galaxy mass and star-formation rate at a fixed $z$.
Combining this with the Kennicutt-Schmidt law, we find
\beq
\label{eq:SK-global}
\mgas(\psi,z) = 2.8 \times 10^{9} \msol 
 \ (1+z)^{-\beta}
 \ \pfrac{\psi}{1 \ \msol \ \rm yr^{-1}}^{\omega}
\eeq
where 
$\beta = 2(1-1/x) = 0.571$ and $\omega = 1/x = 0.714$.
In our model, normal galaxies extend to 
\beq
\label{eq:max-sfr}
\psi \le \psi_{\rm normal}^{\rm max}(z) \equiv \pi r_{\rm disk}(z)^2  \Sigma_{\star,\rm normal}^{\rm max}
 \simeq \frac{450}{(1+z)^{2}} \msol/\rm yr
\eeq
this cutoff becomes
important for $z \ga 1$.

Combining eqs.~(\ref{eq:SK-global}) and~(\ref{eq:gammalum}),
we can express a galaxy's gamma-ray luminosity in terms
of its SFR and redshift:
\beq
\label{eq:lum-psi}
L_\gamma(\psi,z) \propto  \ (1+z)^{-\beta}
 \ \pfrac{\psi}{1 \ \msol \ \rm yr^{-1}}^{\omega+1}
\eeq
Available {\em Fermi} data on resolved $z\approx 0$ star-forming galaxies
are consistent with this scaling \citep{sfgal}.

\subsection{Gamma-Ray Luminosity Function}

A galaxy's luminosity in the \halpha\ line provides a well-established
tracer of star formation rate:
$\lha/1.26 \times 10^{34} \ {\rm W} 
= \psi/1 \ \msol \, {\rm yr^{-1}}$ \citep{hopkins}.  
The \halpha\ distribution of galaxies (luminosity function)
is related to the
cosmic star-formation rate density
via $\csfr(z) = \avg{\psi n} \propto \int \lha \ n(\lha,z) \ d\lha$,
which by now is well-measured as a function of redshift
both by \halpha\ and by other methods, so that its
shape is well-determined observationally.

The gamma-ray luminosity density is a different moment of the
\halpha\ luminosity function
$\grld(z) = \avg{\psi n} \propto (1+z)^{-\beta} \int^{\lha^{\rm max}} \lha^{\omega} \ n(\lha,z) \ d\lha$
via our scaling laws above.
We see that the gamma-ray luminosity distribution (i.e., 
luminosity function) at a given redshift  follows
directly from the distribution of star-formation rates,
as traced by \halpha.  
Since we consider here only normal galaxies, we include only galaxies
with
$\psi \le \psi_{\rm normal}^{\rm max}$ (eq.~\ref{eq:max-sfr})
which sets corresponding limits
$\lha \le \lha^{\rm max}$,
and $L_\gamma \le L_\gamma^{\rm max}$ in eq.~(\ref{eq:grld}).

Current data on the \halpha\ luminosity function can be fit to
a Schechter function, of the form
$n(\tracer,z) \ d\tracer 
 = n_* \ (\tracer/\tracer_*)^{-\alpha} \
 \exp\left( - \tracer/\tracer_* \right)  \
 d\tracer/\tracer_*$
where  $\tracer = \lha$.
Present data are consistent with the value $\alpha = 1.43$,
found for  $z=0$,
persisting for all redshifts.
Data also fix the $z=0$ values
$n_*(0) = 1.0 \times 10^{-3} \ {\rm Mpc^{-3}}$
and 
$\tracer_*(0) =   9.5 \times 10^{34} \ \rm W$,
for $h = 0.71$
\citep{nakamura}. 
This corresponds to
a star-formation rate $\psi(\tracer_*) = 7.5 \ \msol/{\rm yr}$.

However, observations currently do not give unambiguous solutions
for the other two parameters, the characteristic
comoving density of star-forming galaxies $n_*$ and 
the characteristic \halpha\ luminosity $\tracer_*$
\citep{hopkins}.
Moreover, it is unclear whether and  how each parameter
{\em evolves} with redshift.

Two limiting cases bracket the possible behaviors of the
\halpha\ luminosity function and thus of cosmic star formation.
In the case of {\em pure luminosity evolution}
the comoving density of stars is fixed, $n_* = const$ independent of $z$,
and all redshift evolution lies in $\tracer_* = \tracer_*(z)$.
Conversely, {\em pure density evolution}
places the redshift evolution in the density scale
$n_*(z)$ while setting $\tracer_* = const$.
In each of these two limits, we can find the redshift dependence of
the parameters via the requirement
$\avg{L n} \propto \csfr$.  
The redshift history of cosmic-star formation is well-known \citep{horiuchi},
and we encode this in the dimensionless ``shape'' function
\beq
S(z) \equiv \csfr(z)/\csfr(0) . \
\eeq
We then have 
$\tracer_*(z)/\tracer_*(0) = S(z)$ in the case of pure luminosity 
evolution, 
and $n_*(z)/n_*(0) = S(z)$ in the case of pure density evolution.

At a given redshift, eq.~(\ref{eq:mavg}) gives the scaling
$\avg{\mgas} = \avg{\mgas \psi n}/\csfr \propto \avg{\mgas \psi n}/S(z)$.
For our \halpha\ luminosity function and Kennicutt-Schmidt relation we find
a local value of $\avg{\mgas}_{z=0} =6.8 \times 10^{9} \ \msol$.
At other redshifts we have
$\avg{\mgas \psi n} \propto (1+z)^{-\beta} \tracer_*(z)^\omega S(z)$,
and so 
$\avg{\mgas} \propto (1+z)^{-\beta} \tracer_*(z)^\omega$.
Thus for the pure luminosity case $\tracer_* \propto S(z)$,
we find that the gas mass strongly evolves
as $\avg{\mgas} \propto (1+z)^{-\beta} S(z)^\omega$, 
in response to the strongly changing SFR.  
Consequently, the factor of $10$ rise in cosmic star-formation
at $z \simeq 1$ implies a net gamma-ray luminosity increase of a factor $\simeq 30$.
On the other hand, in the pure density evolution
case, galaxy SFRs are constant, $\tracer_*(z) = const$,
so that the mean gas mass
$\avg{\mgas} \propto (1+z)^{-\beta}$ actually {\em decreases}
with redshift, while the comoving number of star-forming galaxies
increases, but the net enhancement at high
redshift is smaller than in the pure luminosity evolution case.
This key difference leads to the factor $\sim 4$ between
the EGB predictions for the pure density and pure luminosity evolution cases seen
in Figure~1.

\section{Results and Conclusions}

Our full numerical calculation
uses a Milky Way pionic source spectrum
whose {\em shape} is derived from
\citet{pfrommer}, calibrated to observations by
normalizing the $> 100$ MeV photon emission per hydrogen atom to
the {\em Fermi} result at intermediate Galactic latitudes
\citep{FermiMW09}.
The cosmic SFR is from ~\citet{horiuchi}.
For the Milky Way SFR, used to normalize the cosmic-ray flux/SFR ratio, we use the recent estimate of \citet{robitaille} ($\psi_{\rm MW} = 1 {M_\odot/\rm yr}$, a factor of 3 lower than earlier work).

Figure~1 shows our results for the normal galaxy contribution to the EGB.
We plot predictions for the limiting cases of pure luminosity and of pure density 
evolution.
The uncertainties in the model inputs, summed in quadrature, propagate  into the displayed error band 
that applies to each curve, which we estimate to be a factor of 
$10^{\pm 0.3}$, resulting from uncertainties of:
30\%  in pionic emissivity \citep{FermiMW09},
40\%   in the normalization of the Galactic
star-formation rate \citep{robitaille},
40\% in the cosmic star-formation rates
\citep{horiuchi}, and 25\% in
the luminosity scaling in eq.~(\ref{eq:mavg}).  The true systematic uncertainty would also reflect
the idealizations in our model (universal cosmic-ray spectra and confinement).
These errors are hard to estimate but in any case imply that the 
uncertainty range in Figure~1 is a lower bound to the error budget.

Within errors, our predictions for both limiting models fall at or 
below the level of the {\em Fermi} data, where the data seem to support the pure luminosity evolution case that
explains nearly the entire signal.   Comparing central values, this model gives $\approx 50\%$ of the
{\em Fermi} EGB $\la 10$ GeV. Thus, unresolved
normal galaxies
make a substantial and likely dominant contribution to the observed EGB,
without overpredicting the signal.
Even the pure density
evolution case accounts for a minimum of 20\% of the EGB
around 0.3 GeV; this provides a {\em lower} limit to the normal-galaxy signal.
Thus, any {\em other} EGB sources \citep{ss96,dermer07,fermicounts,mspulsars}
must contribute no more than the remaining
$80\%$ of the data.
Indeed, the LAT team {\em upper} limit to the blazar EGB contribution 
shown in Figure~1 is comparable to our {\em lower} limit
\citep{fermicounts}.  

The spectral shapes of the 
two limiting cases are very similar:
the peak in $\eobs^2 dI/d\eobs$ lies at $\sim 0.3$ GeV 
because the bulk of the signal comes from $z \sim 1$.  These models predict that the EGB 
turns over for $\eobs \la 0.3$ GeV,  a testable prediction of our model.
For hadronic emission, the high-energy spectral index is the same
as the underlying proton spectral index, here $s_{\gamma,\rm had} = s_{\rm cr} = 2.75$;
this is somewhat steeper than the {\em Fermi} single-power-law fit $s_{\rm obs} = 2.41 \pm 0.05$.
Consequently, our predictions at high energies ($\ga 10$ GeV) fall below the data.
If normal galaxies had a {\em distribution}
of cosmic-ray spectral indices, 
the resulting EGB spectrum
would steepen at high energies where 
the hardest sources
would dominate,
developing a {\em convex tail}. 
Indeed, the {\em Fermi} EGB data suggests a slight 
flattening of slope around $E \ga 10$ GeV,
which might hint at such a transition. 

A galaxy with characteristic \halpha\
luminosity $\tracer_*$ has 
$L_\gamma^*(>100 {\rm MeV}) = 1.4 \times 10^{43} \ {\rm s^{-1}}$.
Such objects have flux $F$ if they lie at distances
$r_* = (L_\gamma^*/4\pi F)^{1/2} = 
11 \, {\rm \ Mpc} \ ({10^{-9} \ \rm cm^{-2} \ s^{-1}}/F)^{1/2}$.
Thus {\em Fermi} should eventually resolve 
\beq
N(>F) \sim 4\pi r_*^3 \ n_*(0)/3
 = 5 \ \pfrac{{10^{-9} \ \rm cm^{-2} \ s^{-1}}}{F}^{3/2}
\eeq
normal galaxies, 
consistent with 2--3 detections to date 
\citep[the LMC, SMC, and perhaps M31;][]{fermi-LMC,fermi-SMC,
fermi-M31}.

Our results do not account for starburst galaxies,
nor for inverse-Compton emission from any
star-forming galaxies;
these  must contribute to the star-forming EGB,
and could have hard spectra dominating $\ga 10$ GeV.
We have also neglected  gamma-ray attenuation by
extragalactic background light  
\citep[important at $E \ga 30$ GeV; e.g.,][and references therein]{sms}.
We will address these issues in future work.

\begin{figure}
\epsscale{.80}
\includegraphics[angle=-90, width=\linewidth]{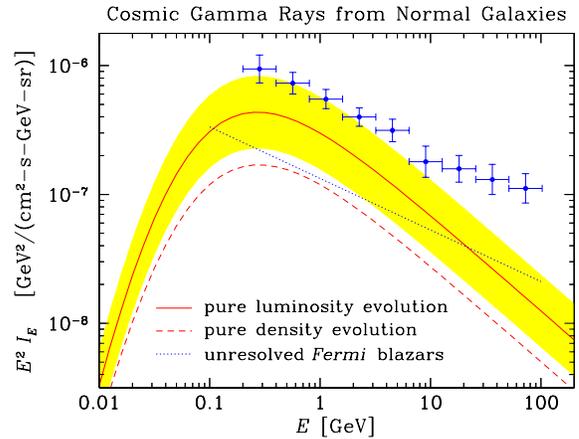}
\caption{
The normal galaxy contribution to the 
extragalactic gamma-ray background.
Curves 
represent two limiting cases of cosmic star formation:  pure luminosity 
and pure density evolution.  
Colored error band illustrates the factor $10^{\pm 0.3}$ uncertainty
in the normalization of both theory curves.
{\em Fermi} data are from \citep{fermi-EGB}.
Dotted line:   unresolved blazar
EGB upper limit
\citep{fermicounts}. 
}
\end{figure}

The amplitude and configuration of magnetic fields in a galaxy have an additional effect on the scaling of cosmic-ray flux with SFR. Confirmation that 
normal galaxies comprise the bulk of the Fermi signal would constitute a unique probe of the evolution of these magnetic fields between the redshift of peak star formation and today. 

Because of their ubiquity, normal galaxies
produce the smallest anisotropies in the EGB, far less than
blazars or other proposed sources.  Thus, by studying
the observed EGB anisotropy as a function of energy,
it may be possible to disentangle the spectrum and 
amplitude of the normal-galaxy contribution
from that of other sources \citep{vaso09}.
Moreover, because normal galaxies seem to 
dominate the {\em Fermi} EGB,
their small contribution to anisotropies will fortuitously provide
the optimal chance of finding smaller and more exotic sources
in the observed signal \citep{sgp09}.

\acknowledgments

It is a pleasure to thank Marco Ajello, Chuck Dermer, Hai Fu, Troy
Porter, and Andy Strong for stimulating discussions.  BDF 
thanks the Goddard Space Flight Center for  hospitality while
some of this work was done.  This work was partially supported by NASA, via
Fermi GI Program grants NNX09AT74G and NNX09AU01G,
and the Astrophysics Theory Program through award 
NNX10AC86G.
VP acknowledges NASA  support
through Einstein Postdoctoral Fellowship grant number
PF8-90060 awarded by the Chandra X-ray Center, which is operated by
the Smithsonian Astrophysical Observatory for NASA under contract
NAS8-03060.  VP would like to thank the
Physics Department at the University of Crete for their hospitality
during part of this work.  The work of TP is supported in part by the
Provincial Secretariat for Science and Technological Development, and
by the Ministry of Science of the Republic of Serbia under project
141002B.

\clearpage

\end{document}